\documentstyle[emulateapj]{article}

\begin{document}

\def\gsim{{{}_>\atop{}^{{}^\sim}}}
\def\lsim{{{}_<\atop{}^{{}^\sim}}}
\def\cross{\rm cross}
\def\kms{{\rm km}\,{\rm s}^{-1}}
\def\kpc{{\rm kpc}}
\def\min{{\rm min}}
\def\max{{\rm max}}
\def\lim{{\rm lim}}
\def\dls{{D_{\rm LS}}}
\def\dos{{D_{\rm S}}}
\def\dol{{D_{\rm L}}}
\def\te{{t_{\rm E}}}
\def\thetae{{\theta_{\rm E}}}
\def\re{{r_{\rm E}}}
\def\b{{\rm base}}
\def\ur{{u_{\rm r}}}
\def\tep{{t_{\rm E}^{\perp}}}
\def\bu{{\bf u}}
\def\cc{{\rm cc}}
\def\bmu{\hbox{$\mu\hskip-7.5pt\mu$}} 
\def\cm{{\rm cm}}
\def\rel{{\rm rel}}
\def\br{{\bf r}}
\def\bv{{\bf v}}
\def\bd{{\bf d}}
\def\e{{\rm E}}
\def\days{\rm days}

\lefthead{ALBROW ET AL.}
\righthead{LIMITS ON GALACTIC PLANETS}

\title{Limits on the Abundance 
of Galactic Planets From\\ Five Years of PLANET Observations}

\author{
M. D. Albrow\altaffilmark{1,2}, J. An\altaffilmark{3},
J.-P. Beaulieu\altaffilmark{4}, J. A. R. Caldwell\altaffilmark{5},
D. L. DePoy\altaffilmark{3}, M. Dominik\altaffilmark{6},
B. S. Gaudi\altaffilmark{3,7,8},
A. Gould\altaffilmark{3},J. Greenhill\altaffilmark{9}, K. Hill\altaffilmark{9},
S. Kane\altaffilmark{2,9}, R. Martin\altaffilmark{10},
J. Menzies\altaffilmark{5}, R. M. Naber\altaffilmark{6},
J.-W. Pel\altaffilmark{6},R. W. Pogge\altaffilmark{3},
K. R. Pollard\altaffilmark{1}, P. D. Sackett\altaffilmark{6},
K. C. Sahu\altaffilmark{2},P. Vermaak\altaffilmark{5},
P. M. Vreeswijk\altaffilmark{6,11}, R. Watson\altaffilmark{9},
A. Williams\altaffilmark{10} \\
The PLANET Collaboration}

\altaffiltext{1}{Univ.\ of Canterbury, Dept.\ of Physics \& Astronomy, 
Private Bag 4800, Christchurch, New Zealand}
\altaffiltext{2}{Space Telescope Science Institute, 3700 San Martin Drive, 
Baltimore, MD. 21218~~U.S.A.}
\altaffiltext{3}{Ohio State University, Department of Astronomy, Columbus, 
OH 43210, U.S.A.}
\altaffiltext{4}{Institut d'Astrophysique de Paris, INSU CNRS, 98 bis 
Boulevard Arago, F-75014, Paris, France}
\altaffiltext{5}{South African Astronomical Observatory, P.O. Box 9, 
Observatory 7935, South Africa}
\altaffiltext{6}{Kapteyn Astronomical Institute, Postbus 800, 
9700 AV Groningen, The Netherlands}
\altaffiltext{7}{Institute for Advanced Study, Einstein Drive, Princeton, NJ 08540}
\altaffiltext{8}{Hubble Fellow}
\altaffiltext{9}{Univ. of Tasmania, Physics Dept., G.P.O. 252C, 
Hobart, Tasmania~~7001, Australia}
\altaffiltext{10}{Perth Observatory, Walnut Road, Bickley, Perth~~6076, Australia}
\altaffiltext{11}{Astronomical Institute ``Anton Pannekoek'', University of 
Amsterdam, Kruislaan 403, 1098 SJ Amsterdam, The Netherlands}

\begin{abstract}
	We search for signatures of planets in 43 intensively
monitored microlensing events that were observed between 1995 and
1999.  Planets would be expected to cause a short duration ($\sim 1\,{\rm
day}$) deviation on the smooth, symmetric light curve
produced by a single-lens.  We find no such anomalies and infer that
less than 1/3 of the $\sim 0.3\,M_\odot$ stars that typically comprise
the lens population have Jupiter-mass companions with semi-major axes
in the range of $1.5\,{\rm AU}<a<4\,{\rm AU}$.  Since orbital periods
of planets at these radii are 3-15 years, the outer portion of this
region is currently difficult to probe with any other technique.

\end{abstract}


\section{Introduction}

	Searches for extrasolar planets are being carried out using
several methods.  More than 60 planets have been discovered by the
doppler-shift technique (Marcy, Cochran, \& Mayor 2000).  While
ground-based astrometric searches have not yielded any definitive
detections, future astrometric satellites
are expected to radically improve the sensitivity of this technique
(Lattanzi et al.\ 2000).  The occultation method has yielded an
important null result for planets in 47 Tuc (Gilliland et al.\ 2000), and
one confirmation (Charbonneau et al.\ 2000; Henry et al.\ 2000).  
All of these methods are fundamentally
restricted to planets with orbital times shorter than the experiment,
and hence to relatively close (and also generally massive) companions.

	Microlensing provides a method to search for planets (Mao \&
Paczy\'nski 1991) that does not suffer from this limitation.  When two
stars are approximately aligned with an observer, the nearer star (the
``lens'') splits the light from the more distant star (the ``source'')
into two images whose brightnesses change as the
relative alignment changes.  The characteristic angular scale,
$\theta_\e$ (``Einstein ring''), and timescale, $t_\e$, of such a
microlensing event are
\begin{equation}
\theta_\e = \sqrt{{4 G\over c^2}\,{M\over D_\rel}},\qquad t_\e = 
{\theta_\e\over \mu_\rel},
\label{eqn:thetaedef}
\end{equation}
where $M$ is the mass of the lens, $D_\rel\equiv {\rm AU}/\pi_\rel$,
$\pi_\rel$ is the lens-source relative parallax, and $\mu_\rel$ is the
relative proper motion.  Note that $D_\rel^{-1}=D_L^{-1}-D_S^{-1}$,
where $D_L$ and $D_S$ are the lens and source distances.  For typical
events seen toward the Galactic bulge, $\theta_\e\sim 320\mu{\rm
as}(M/0.3\,M_\odot)^{1/2}$, which is too small to resolve
directly.  However, since the images are magnified, the event can be
identified photometrically.  For typical bulge events, $\mu_\rel\sim
25\,\kms\,\kpc^{-1}$, so $t_\e\sim 20\,$days, and hence nightly
monitoring is sufficient to find most events.  Four groups, OGLE,
MACHO, EROS, and MOA have carried out such microlensing searches and
combined have detected over 700 events, most in the direction of the
Galactic bulge (Udalski et al.\ 2000; Alcock et al.\ 1997; 
Abe et al.\ 1997).  All four teams 
recognize these events in real time and electronically alert the
community soon after the onset of the event.

	If the lens has a planet that lies close to one of the lensed
images of the source, that image is further perturbed and the
magnification changes significantly during a time $t_p$,
\begin{equation}
t_p = {\theta_p\over\theta_\e}t_\e,\qquad \theta_p =\sqrt{m_p\over
M}\theta_\e,
\label{eqn:thetapdef}
\end{equation}
where $m_p$ is the mass of the planet.  Hence a planet betrays itself
as a short ($\sim 1\,$day) ``bump'' on an otherwise normal single-lens
light curve (Refsdal 1964; Paczy\'nski 1986).  
Gould \& Loeb (1992) showed that, with photometric precision of about 1-2\%,
Jupiter-mass planets present a reasonable probability for detection
throughout a wide zone centered on the Einstein ring.  For typical
lens distances $\sim 6\,\kpc$, this sensitivity peaks at projected
separations $2\,{\rm AU}(M/0.3\, M_\odot)^{1/2}$.  Since $t_p$ is of
order or shorter than the sampling time of the microlensing search
teams, Gould \& Loeb (1992) advocated setting up a globe-straddling
network of observatories to do continuous follow-up observations of
alerted events.

	The PLANET collaboration was formed in 1995 (Albrow et al.\
1996) expressly to carry out such observations and demonstrated in
that pilot year that the program was feasible (Albrow et al.\ 1998).
PLANET obtained substantial observing time at four southern locations
(Tasmania, Western Australia, South Africa, and Chile) during 1995-99.  
During these five years, we monitored
$\sim 50$ microlensing events sufficiently well to have good to
excellent sensitivity to planets, none of which displayed a clear
photometric anomaly that was best explained by a planet orbiting
the lens.  We quantify this statement by characterizing the
statistical sensitivity of our five-year data set to planets.  We then
use these results to build mass-separation exclusion diagrams for the
typical Galactic stars (i.e., microlenses) that our survey probes.
The basic method of analysis is given in Gaudi \& Sackett (2000) and
applied to event OGLE~1998-BUL-14 in Albrow et al.\ (2000b). The details of its application 
to the present data set are given by Albrow et al.\ (2001).

\section{PLANET Five-Year Photometric Dataset}

	Our data were acquired over five years from six telescopes:
the Canopus 1m near Hobart, Tasmania, the Perth/Lowell 0.6m at
Bickley, Australia, the Elizabeth 1m at the South African Astronomical
Observatory (SAAO) at Sutherland, South Africa, the ESO/Dutch 0.9m at
La Silla, Chile, the Yale 1m and the CTIO 0.9m at the Cerro Tololo
Interamerican Observatory (CTIO) at La Serena, Chile.  
Data were collected in Cousins $I$ and
Johnson $V$, with strong emphasis on the former.

The bulk of events analyzed here have median sampling times of
1 to 2 hours, or ${\cal O}(10^{-3}\te)$.  
Observations from four observatories permit round-the-clock monitoring of events;
the best events rarely have gaps in the data longer than a day. 
For such events, the typical photometric precision (as judged by the scatter) is of order
1--2\% for points near the peak, which contain most of the sensitivity
to planets.  The dataset contains 5-6 high-magnification events
with well-sampled peaks. These events contribute at least 1/2 of our overall sensitivity. 
For the final event sample,  the median number of photometric points within
$\te$ of the peak magnification is about 140, with $75\%$ of events
having more than 65 points and $25\%$ having more than 250 points
within $\te$ of the peak.  Details are given in Albrow et al.\ (2001).

\section{Event Selection}

We begin with the complete sample of Galactic bulge events monitored
by PLANET during 1995-1999, discarding those of extremely poor
quality and those known to contain
anomalies characteristic of roughly equal-mass binaries or other
anomalies unrelated to binarity.  Normal
point-source/point-lens (PSPL) microlensing events are described by
$$
F(t) = F_s A(t) + F_b,
$$
$$
A[u(t)]= {u^2+2\over u(u^2+4)^{1/2}},
$$
\begin{equation}
\qquad u(t) = \sqrt{u_0^2 + {(t-t_0)^2\over t_\e^2}},
\label{eqn:foft}
\end{equation}
where $F$ is the observed flux, $F_s$ is the source flux, $F_b$ is the flux from any
background light that is not magnified, $u_0$ is the projected source-lens impact parameter,
and $t_0$ is the time of closest approach.  Separate $F_s$ and $F_b$
are required for each observatory and wave band.

Non-planetary light curve anomalies are identified
through explicit modeling. We discard such anomalous light curves from
our analysis because we do not currently have the ability to
systematically search for planetary signatures in the presence of
these secondary effects.  We note, however, that none of the excluded
light curves show signs of short duration bumps, except
MACHO~97-BLG-41 for which the deviation is explained naturally by
binary rotation (Albrow et al.\ 2000a) rather than a planet orbiting a
binary (Bennett et al.\ 1999).

To eliminate events of particularly poor quality for planet detection,
we introduce event selection criteria:
\hfil\break\noindent{1)} All data must pass certain quality tests:
only DoPhot (Schechter, Mateo, \& Saha 1993) types 11 and 13 are accepted.
\hfil\break\noindent{2)} There must be at 
least 20 data points from at least one observatory in one band.
\hfil\break\noindent{3)} Each observatory-band dataset included
must contain at least 10 data points.
\hfil\break\noindent{4)} The error in $u_0$ from a combined fit must be less 
than 50\%.
\hfil\break\noindent
If $u_0$ is not well constrained, then the source's path through the
Einstein ring is not well determined, and hence it is difficult to
estimate the sensitivity of the event to planets.  When available, we
use OGLE and MACHO data to help constrain $u_0$, but not to search for
planets.  Our final sample includes a total of 43 events.  
A full list of these events along with the photometric data 
is presented in Albrow et al.\ (2001).

\section{Systematic Search for Planetary Signatures}

	The method of measuring the sensitivity of an event to the
presence of planets, and at the same time searching for planetary
signatures if they are present, is thoroughly described in Albrow et
al.\ (2001).  Very briefly, we obtain a PSPL fit using equation
(\ref{eqn:foft}), and simultaneously renormalize the photometric
errors at each observatory so that the total $\chi^2_{\rm PSPL}$ of
this fit is equal to the number of degrees of freedom.  Note that
outliers are not included for error renormalization, but are included
when searching for planets.  
We also include in all model fits a term that accounts for
the correlation of the photometry with the seeing that we observe in
most of our light curves (both microlensed and constant stars).
 Although these  procedures do bias us against
binaries, the bias is only serious if all the points from one observatory are 
concentrated in a short span of the light curve and there are no contemporaneous 
data from other observatories.  From direct inspection of the
43 lightcurves in our event sample we find that such bunching affects ${\cal O}(1\%)$ of our
total lightcurve coverage, which leads to an overestimate of our detection efficiency of a
similar magnitude.  Since this bias is an order of magnitude smaller than our statistical errors, 
we ignore it.  On the other hand, Monte Carlo experiments with constant stars reveal that 
error renormalization and removal of systematic effects are essential in
order to draw reliable inferences from the light curves and to avoid spurious detections.  
See Albrow et al.\ (2001) for details and a thorough discussion.

For each
planet-star mass ratio $q$ and each planet-star projected separation
$\theta_\e d$, as well as each angle $\alpha$ of the source trajectory
relative to the binary axis, we find the best fit to the remaining
parameters $(t_0,t_\e,u_0, F_s,F_b)$.  The corresponding $\chi^2$ thus
yields $\Delta\chi^2\equiv\chi^2-\chi^2_{\rm PSPL}$, for which we set
a threshold value $\Delta\chi^2_\min=60$.  If
$\Delta\chi^2>\Delta\chi^2_\min$, then the geometry $(d,q,\alpha)$ is
excluded.  If $\Delta\chi^2<-\Delta\chi^2_\min$, we tentatively
conclude that we have detected a planet.  The detection efficiency
$\epsilon_i(d,q)$ for planets with this $(d,q)$ in event $i$ is then
just the fraction of all angles $\alpha$ (out of $2\pi$) that are
excluded.  We define a ``planetary system'' as a binary lens with mass
ratio $q<0.01$.

\centerline{{\vbox{\epsfxsize=9.0cm\epsfbox{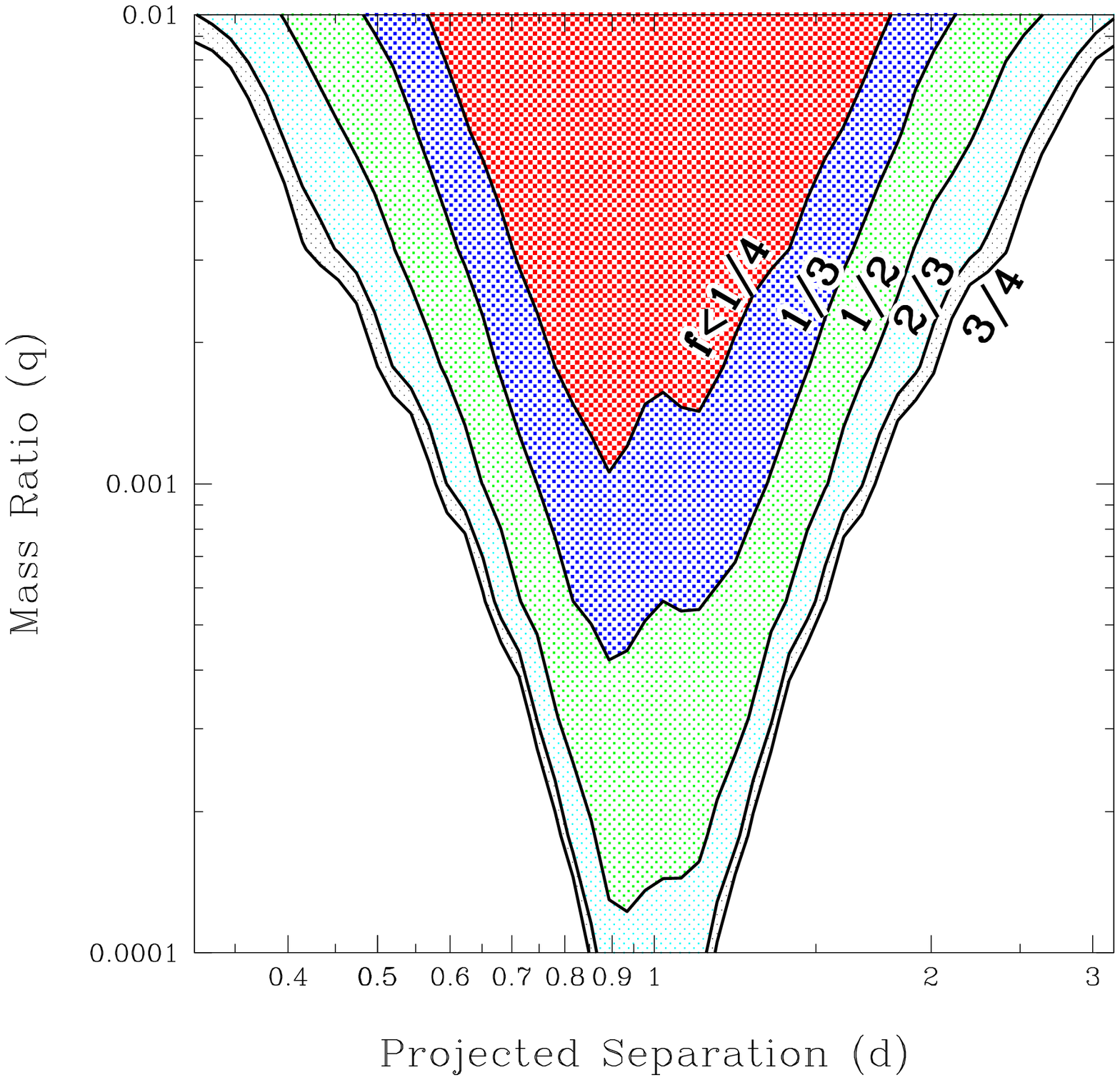}}}}
{\footnotesize {\bf FIG. 1}
Exclusion diagram for pairs of planet parameters $(d,q)$, where $d$ is
the projected separation in units of the Einstein ring, and $q$ is the
planet-star mass ratio.  The inner contour indicates the $(d,q)$ for
which the fraction of lenses with a planet is $f<1/4$ at 95\%
confidence.  Other contours are for $f<1/3$, 1/2, 2/3, and 3/4.  A
mass ratio $q=0.001$ corresponds approximately to $m_p=0.3\,M_{\rm
Jup}$, and a projected separation of $d=1$ corresponds to a physical
projected separation $r_p=2\,$AU.  We have assumed a detection
threshold of $\Delta\chi^2_\min = 60$.  }
\bigskip

	We set the threshold $\Delta\chi^2_\min=60$ by first noting
the continuous distribution of $\Delta\chi^2\la 50$ in our data.  If a
significant fraction of these were due to planets, the distribution
would extend to much more extreme values, since a small random change
in the impact parameter could easily increase $\Delta\chi^2$ to
several hundred.  Hence, the great majority of these deviations must
be due to small unrecognized systematic errors.  Monte Carlo tests 
performed with constant stars reveals that deviations of $\Delta\chi^2 \la 60$ are
easily explained by systematic and statistical noise. We therefore set the
threshold high enough to exclude these non-planetary sources of
noise.  We also show results for the more conservative
threshold $\Delta\chi^2_\min=100$.

\centerline{{\vbox{\epsfxsize=9.0cm\epsfbox{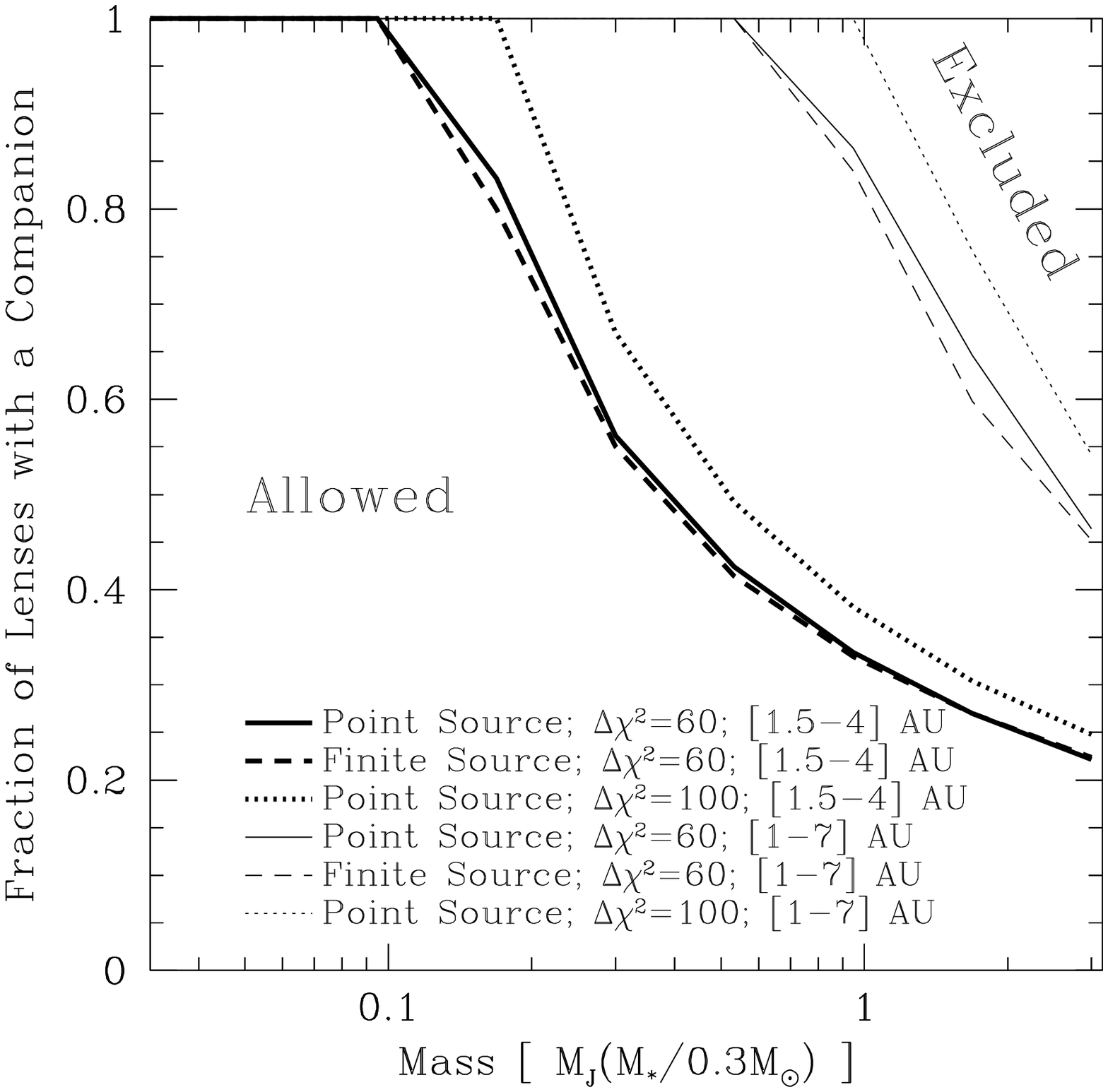}}}}
{\footnotesize {\bf FIG. 2}
Exclusion diagram for planets anywhere in a continuous range of
semi-major axes centered on the Einstein ring.  Bold curves show the
excluded fraction (at 95\% confidence) anywhere in the range
$1.5\,{\rm AU}<a<4\,$AU, while solid curves show the fraction for
$1\,{\rm AU}<a<7\,$AU. We have assumed a detection threshold of
$\Delta\chi^2_\min = 60$, and that the primary has an Einstein ring
radius $\re=2~{\rm AU}$.  For other Einstein ring radii, the
separations scale as $(\re/2~{\rm AU})$. The dashed curves show the
negligible effect of including the finite size of the source in the
modeling.  The dotted lines are for $\Delta\chi^2_\min=100$.  }
\bigskip

This search procedure identifies two possible candidates,
but these are also representatives of two classes of phenomena that
must be excluded from our search: nearly equal-mass binary lenses and
global-asymmetry anomalies.  MACHO~99-BLG-18 has a $\sim 15\,$day
anomaly of amplitude $\sim 2\%$.  Such an anomaly is longer than that
expected from planets with $q\la 0.01$, and we therefore
systematically explored binary-lens fits with $q \ge 0.01$.  This
uncovered a fit with $q\sim 0.2$ that is favored over the best-fit
planet ($q \le 0.01$) by $\Delta\chi^2=22$.  We therefore exclude
MACHO~99-BLG-18 from the analysis. Although $\Delta\chi^2=22$ is below our normal threshold 
($\Delta\chi^2=60$), we estimate that the probability that we have inadvertently
thrown out a real planetary detection that is $\la 10\%$, 
smaller than the statistical errors on our
resultant limit on planetary companions. See Albrow et~al.\ (2001) for a discussion.
 OGLE~99-BUL-36 displays an overall
asymmetry that is consistent with a distortion caused by a $q\sim
0.003$ planet.  Such parallax asymmetries are a general feature of parallax which must
be present at some level in all microlensing events (Gould, Miralde-Escud\'e \& Bahcall 1994).  
Indeed, this distortion is equally well-fit by 
such a parallax asymmetry model.
Given that OGLE~99-BUL-36 has a relatively short timescale ($\te \simeq 30~\days$), 
one might naively expect the parallax interpretation to be unlikely.
However, the magnitude of the parallax asymmetry is quite small, 
and only detectable due to the high quality of the data.  The resulting asymmetry implies
reasonable values for the most probable mass and distance to the lens.
We conclude that we cannot reliably detect planets from global asymmetries, and should
exclude from our analysis all events that display such anomalies and reduce our efficiency
estimates accordingly.  
Although we do not explicitly do this, this
results in detection efficiencies that are overestimated by a negligible amount,
since a very special planetary geometry is required
to produce a global asymmetry (rather than a short duration anomaly).  This conclusion
is borne out by explicit simulations (Albrow et~al.\ 2001).
Thus there are no viable planet candidates out of our original sample of 43 events.

\section{Exclusion Diagram for Galactic Planets}

       From the above analysis of each event $i$, we obtain an efficiency
$\epsilon_i(d,q)$ as a function of planetary geometry $(d,q)$.  Let
$f(d,q)$  be the fraction of lenses having a planet at $(d,q)$. Then,
from binomial statistics, the probability of observing no planets
is $P=1-\prod_i[1- f(d,q)\epsilon_i(d,q)]$.  Note that in the limit
 $f\epsilon\ll 1$ (approximately valid in the present study), this
reduces to the Poisson formula $P=[1-\exp(-N)]$, where
$N(d,q)=f(d,q)\sum_i \epsilon_i(d,q)$ is the expected number of
detections.
Thus, to a good approximation, fractions $f(d,q)\geq 3/\sum_i
\epsilon_i(d,q)$ can be rejected at 95\% confidence.

Based on this analysis, we build an exclusion diagram (Fig.\
1) based on the sample of 42 events (excluding MACHO-99-BLG-18)
for planet parameters $(d,q)$.  To convert $d$ and $q$
into physical parameters of planet mass $m_p$ and projected physical
separation $r_p$, we must estimate the typical mass $M$ and physical
Einstein radius $D_L\theta_\e$ for the events in our sample.  The
majority of detected microlensing events are almost certainly bulge
stars lensing other bulge stars (Kiraga \& Paczy\'nski 1994).  If the
lenses were drawn randomly from the bulge mass function as measured by
Zoccali et al.\ (2000), one would expect the typical mass to be $M\sim
0.3\,M_\odot$ and hence the typical timescale to be $t_\e\sim
20\,$days.  However, the median timescale for our sample is 40 days.
The difference is probably mainly due to a bias in our selection
process.  From equation
(\ref{eqn:thetaedef}), a bias toward large $t_\e$ will cause biases
toward higher $M$, higher $\pi_\rel$, and lower $\mu_\rel$.  Comparing
Figures 1a and 1b from Gould (2000), we infer that most of the
dispersion in observed timescales is due to $\mu_\rel$ and $\pi_\rel$,
so the bias in terms of mass is likely to be modest.  Hence, we adopt
$M\sim 0.3\,M_\odot$ and $\pi_\rel\sim 40\,\mu$as, and thus
$\theta_\e\sim 320\,\mu$as.  With our convention $D_L=6\,\kpc$,
$q=0.001\Rightarrow m_p=0.3\,M_{\rm Jup}$, and $d=1\Rightarrow
r_p=2\,$AU.

	In Figure 2 we present upper limits for the
fraction of lenses with planets over two ranges of semi-major axes $a$
centered on the Einstein ring.   To convert from projected separation (Fig.\
1) to semi-major axis, we integrate over all orientations
assuming circular orbits, $M=0.3\,M_\odot$ and $D_L\theta_\e=2\,$AU.
We calculate efficiencies using both the point-source approximation
and allowing for finite source size (Gaudi \& Sackett 2000) but find
that the difference is negligible.  We find that less than 1/3 of lenses
have Jupiter-mass companions anywhere in the range of $1.5\,{\rm
AU}<a<4.0\,$AU.   These are the first significant
limits on planetary companions of M dwarfs.

\begin{acknowledgements}

We thank the anonymous referee for helping to clarify a number of important points. 
We thank the EROS, MACHO, and OGLE collaboration for providing the
original electronic microlensing event alerts that are the sine qua
non of this work, and the MACHO and OGLE collaborations for providing
their original data for many of the events.  We are especially
grateful to the observatories that support our science (Canopus, CTIO,
ESO, Perth, SAAO, Yale) via the generous allocations of time that make
this work possible, and to those who have donated their time to
observe for this program.   
PLANET acknowledges financial support via
award GBE~614-21-009 from the organization for {\sl Nederlands
Wetenschappelijk Onderzoek\/} (Dutch Scientific Research), the Marie
Curie Fellowship ERBFMBICT972457 from the European Union,
grant Deutsche Forschungsgemeinschaft Do 629/1-1, a ``coup de
pouce 1999'' award from the {\sl Minist\`ere de l'\'Education
nationale, de la Recherche et de la Technologie, D\'epartement
Terre-Univers-Environnement\/}, grants AST~97-27520 and AST~95-30619
from the NSF,  NASA grant NAG5-7589, a Presidential Fellowship from the Ohio State University, and NASA through a Hubble Fellowship grant from the Space Telescope Science Institute, which is operated by the Association of Universities for Research in Astronomy, Inc., under NASA contract NAS5-26555.

\end{acknowledgements}

\newpage

\end{document}